\documentclass[prd,twocolumn,eqsecnum]{revtex4}
\usepackage{bm} 
\usepackage{amssymb}
\usepackage{latexsym} 
\newcommand{\base}[2]{e^{#1}_{#2}}
\begin{document}
\title{A reformulation of the Barrab\`es-Israel null-shell formalism}        
\author{Eric Poisson}
\affiliation{Department of Physics, University of Guelph, Guelph, 
Ontario, Canada N1G 2W1}
\date{July 5, 2002}   
\begin{abstract}
We consider a situation in which two metrics are joined at a null
hypersurface. It often occurs that the union of the two metrics gives
rise to a Ricci tensor that contains a term proportional to a Dirac
$\delta$-function supported on the hypersurface. This singularity is 
associated with a thin distribution of matter on the hypersurface, and
following Barrab\`es and Israel, we seek to determine its
stress-energy tensor in terms of the geometric properties of the null
hypersurface. While our treatment here does not deviate strongly from
their previous work, it offers a simplification of the computational
operations involved in a typical application of the formalism, and it
gives rise to a stress-energy tensor that possesses a more
recognizable phenomenology. Our reformulation of the null-shell
formalism makes systematic use of the null generators of the singular
hypersurface, which define a preferred flow to which the flow of
matter can be compared. This construction provides the stress-energy
tensor with a simple characterization in terms of a mass density, a
mass current, and an isotropic pressure. Our reformulation also
involves a family of freely-moving observers that intersect the
surface layer and perform measurements on it. This construction gives
operational meaning to the stress-energy tensor by fixing the argument
of the $\delta$-function to be proper time as measured by these
observers. 
\end{abstract}
\maketitle

\section{Introduction}
\label{I} 

Many applications of the general theory of relativity involve two 
metrics being joined at a common boundary, a hypersurface that
partitions spacetime into two distinct regions. It often occurs that
the union of the metrics does not form a smooth solution to the
Einstein field equation, even when the metric is continuous at the
hypersurface. Indeed, a discontinuity in the metric's transverse
derivative always produces a singularity in the Riemann tensor. The
singularity, however, is sufficiently mild (it is a Dirac
$\delta$-function with support on the hypersurface) that it yields
itself to a sound physical interpretation. 

When the hypersurface is timelike (or spacelike), only the Ricci part
of the Riemann tensor is singular, and the singularity can be
associated with a material stress-energy tensor supported on the
hypersurface. When the hypersurface is null, on the other hand, both
the Weyl and Ricci parts of the Riemann tensor are singular; here the
singularity in the Ricci tensor can just as well be associated with a
material stress-energy tensor, while the singularity in the Weyl
tensor is interpreted as an independent impulsive tidal wave. The 
material present on the hypersurface, which gives rise to a singular
Ricci tensor, is often referred to as a {\it surface layer}, or as a  
{\it thin shell}. The dynamics of a timelike or null surface layer,
whose world surface coincides with the singular hypersurface, can be
of considerable interest and has been the subject of numerous studies.   

The standard formalism to describe the physical properties of a 
timelike (or spacelike) surface layer was produced by Israel more than 
35 years ago \cite{israel}. This practical formalism allows for
efficient computations of the shell's surface stress-energy tensor in
terms of the discontinuity of the extrinsic curvature across the
hypersurface. The Israel formalism has even been implemented as the
package {\tt GRJunction} \cite{musgrave1} within the computer algebra
program {\tt GrTensorII} \cite{grii}.  

Several extensions of the Israel formalism to the case of a null    
hypersurface were proposed in the following years \cite{penrose, taub,
redmount, berezin, clarke, barrabes}, but a definitive treatment had
to wait the 1991 work of Barrab\`es and Israel \cite{BI}, who produced
a formulation that was much better suited to practical
applications. Their formalism was used in several interesting
contexts: to model an exploding white hole \cite{poisson}; to
speculate on the possible creation of expanding universes inside
black holes \cite{frolov}; to calculate the gravitational signal that
accompanies a body's sudden change of multipole moments \cite{BH1,
BH2}, or its sudden deceleration to a complete stop \cite{BH4, BH5};
and to calculate the detection of an impulsive null signal by a system
of neighbouring test particles \cite{BH6, BH7}. A number of additional
applications were considered by Barrab\`es and Hogan in their
extension of the original Barrab\`es-Israel formalism \cite{BH3},
which was also extended to the case of scalar-tensor theories
\cite{bressange1, bressange2} and the Einstein-Cartan theory
\cite{bressange3}. The formalism has been implemented within the
computer package {\tt GRJunction} \cite{musgrave2}.   

Our purpose with this paper is to present a reformulation of the 
Barrab\`es-Israel formalism for null surface layers. With this new 
formulation we hope to achieve both a simplification of the
computational steps involved in a typical application of the
formalism, and a clarification of some of the interpretational aspects
of the formalism. While the work presented here does not deviate
strongly from the original Barrab\`es-Israel treatment, we believe
that it will ease and encourage further explorations of the physics of
null surface layers; as such, our work should constitute a useful
addition to the existing literature. In effect, what we set out to do 
is to make the Barrab\`es-Israel formalism more user-friendly, and
more amenable to straightforward physical interpretation.  

The first key aspect of this reformulation of the null-shell formalism
is our systematic involvement of the null generators of the singular
hypersurface. The generators are used to define a preferred flow on
the hypersurface, to which the flow of matter on the surface layer can
be compared. As we shall see, this construction provides the surface
stress-energy tensor with a simple characterization in terms of a mass
density (which represents a flow of matter along the null generators),
a current density (which represents a flow of matter across
generators), and an isotropic pressure. The surface stress-energy
tensor therefore comes with a clear and meaningful phenomenology.  

The null generators also provide us with a preferred family of
intrinsic coordinates on the hypersurface. These are adapted to
generators, in the sense that two of the coordinates serve to label 
the generators, while the third coordinate is the generator's
parameter. The reformulation makes exclusive use of this family of
intrinsic coordinates. While this represents a restriction of the
freedom granted by the original treatment of Barrab\`es and Israel
(who allowed for an arbitrary system of intrinsic coordinates), we
shall see that this choice greatly simplifies the computational
operations that arise in practical applications of the formalism. A
particularly important aspect of this simplification is that once the
intrinsic coordinates have been constructed, all remaining operations
follow a simple algorithm. We admit that the explicit construction of
the preferred coordinates can be a difficult task, and that in some
cases this might motivate a retreat to the original formulation. In
many applications, however, including the ones considered in this
paper, the intrinsic coordinates are easily obtained and the remaining
operations are entirely straightforward.  

The second key aspect of our reformulation is an attempt to give
operational meaning to the $\delta$-function structure of the surface
layer's stress-energy tensor. In the original treatment, the argument
of the $\delta$-function is left arbitrary, and as a result the
stress-energy tensor acquires an ambiguity: it is defined
up to a multiplicative factor that can vary arbitrarily over the
hypersurface. Our reformulation does not eliminate this ambiguity
(which is an unavoidable feature of null hypersurfaces), but it
characterizes it in a physical, well-motivated way. We introduce, in a
neighbourhood of the hypersurface, a family of freely-moving observers
that intersect the surface layer and perform measurements on it. This 
family is arbitrary, and it is this arbitrariness that now generates
the ambiguity associated with the surface stress-energy tensor. Once a  
specific choice of observers has been made, however, all traces of
ambiguity disappear. The argument of the $\delta$-function is then
fixed to be the observer's proper time, which is set to zero at the
hypersurface. This choice confers a robust operational meaning to the
surface stress-energy tensor.  

Our reformulation of the Barrab\`es-Israel null-shell formalism is
presented in Sec.~\ref{II} of the paper; there we omit all derivations
and present the formalism as a ready-to-use recipe. Derivations of the
main results are contained in Sec.~\ref{III}. In Sec.~\ref{IV} we
examine some issues regarding the parameterization of the null
generators and their impact on the physics of null surface 
layers. Both Secs.~\ref{III} and \ref{IV} can be omitted during a
first reading of the paper; they contain technical material that are
not required in practical applications of the formalism. Finally, in
Sec.~\ref{V} we present four simple applications of the formalism, 
designed to illustrate the methods and highlight some aspects of
the physics of null surface layers.                           

\section{Null-shell formalism} 
\label{II} 

We consider a hypersurface $\Sigma$ that partitions spacetime into two
regions ${\cal M}^\pm$. In ${\cal M}^\pm$ the metric is 
$g_{\alpha\beta}^\pm$ when expressed in coordinates $x^\alpha_\pm$; as
indicated, the coordinate systems on each side of the hypersurface may
be different. We assume that the hypersurface is null, and our
convention is such that ${\cal M}^-$ is in the past of $\Sigma$, and 
${\cal M}^+$ in its future. We assume also that the hypersurface is 
singular, in the sense that in a suitable coordinate system defined in
a local coordinate neighbourhood that includes $\Sigma$, the Riemann
tensor is distributional and contains a term proportional to a Dirac
$\delta$-function supported on the hypersurface. In general, both the
Weyl and Ricci parts of the Riemann tensor are singular. The
singularity in the Ricci tensor can be thought to be produced by a
material surface layer whose spacetime history coincides with
$\Sigma$. We wish to give a characterization of this surface layer in
terms of a surface stress-energy tensor. We shall here have nothing to
say about the singularity in the Weyl tensor and the interesting
physics associated with it; for a discussion we refer the reader to
the relevant literature (for example, see \cite{penrose, taub,
redmount, BI, BH3}).   

\subsection*{Intrinsic coordinates}

Our formulation of the null-shell formalism is based on a preferred
family of coordinate systems that are used to represent events on the  
hypersurface; these coordinates are adapted to the null generators of   
$\Sigma$. Since the generators define a preferred flow on the
hypersurface, and since the preferred coordinates give the simplest
description of the generators, this type of intrinsic coordinate
system will be seen to produce an especially simple description of the
surface stress-energy tensor.      

The preferred intrinsic coordinates are    
\begin{equation}
y^a = (\lambda,\theta^A), 
\label{2.1} 
\end{equation}
where the lower-case latin index $a$ runs over the values $(1,2,3)$,
while the upper-case index $A$ runs over $(2,3)$; thus $y^1 =
\lambda$, $y^2 = \theta^2$, and $y^3 = \theta^3$. We assume that the 
coordinates $y^a$ are the same on both sides of $\Sigma$. We take
$\lambda$ to be an arbitrary parameter on the null generators of the
hypersurface, and we use $\theta^A$ to label the generators. Thus, to 
each generator we assign two coordinates $\theta^A$, and to each event
on a given generator we assign a third coordinate $\lambda$. Changing
$\lambda$ while keeping $\theta^A$ constant produces a displacement
along a single generator; changing $\theta^A$ while keeping $\lambda$
constant produces a displacement across generators. It is possible to
choose $\lambda$ to be an {\it affine} parameter on {\it one side} of
the hypersurface; as we shall see in Sec.~\ref{IV}, however, it is in
general not possible to make $\lambda$ an affine parameter on both
sides of $\Sigma$. The coordinate system of Eq.~(\ref{2.1}) is not
unique, as it leaves the freedom of performing an independent
reparameterization of each generator; we shall explore this freedom in
Sec.~\ref{IV}.  

\subsection*{Tangent vectors} 

As seen from ${\cal M}^\pm$, $\Sigma$ is described by a set of
parametric relations $x^\alpha_\pm(y^a)$, and using these we can
introduce tangent vectors $\base{\alpha}{\pm a} = \partial
x^\alpha_\pm /\partial y^a$ on each side of the hypersurface. These
are naturally segregated into a null vector $k^\alpha_\pm$ that is
tangent to the generators, and two spacelike vectors
$\base{\alpha}{\pm A}$ that point in the directions transverse to the
generators. Explicitly,  
\begin{equation}
k^\alpha = \biggl( \frac{\partial x^\alpha}{\partial \lambda}
\biggr)_{\!\! \theta^A} 
\equiv \base{\alpha}{\lambda}, \qquad 
\base{\alpha}{A} = \biggl( \frac{\partial x^\alpha}{\partial\theta^A}
\biggr)_{\!\! \lambda}.   
\label{2.2} 
\end{equation}
(Here and below, in order to keep the notation simple, we refrain from
using the ``$\pm$'' label in displayed equations; this should not
create any confusion.) By construction, these vectors satisfy    
\begin{equation}
k_\alpha k^\alpha = 0 = 
k_\alpha \base{\alpha}{A}. 
\label{2.3}
\end{equation}
The remaining inner products
\begin{equation}
\sigma_{AB}(\lambda,\theta^C) \equiv 
g_{\alpha\beta}\, \base{\alpha}{A} \base{\beta}{B}
\label{2.4}
\end{equation}
do not vanish, and we assume that they are the same on both sides of   
$\Sigma$:  
\begin{equation}
\bigl[ \sigma_{AB} \bigr] = 0. 
\label{2.5}
\end{equation}
We have introduced here the standard notation $[A] \equiv 
A({\cal M}^+)|_\Sigma - A({\cal M}^-)|_\Sigma$ for any scalar quantity
$A$. In the first term, $A$ is computed in ${\cal M}^+$ and evaluated
on $\Sigma$; in the second term, $A$ is computed in ${\cal M}^-$
instead. A vanishing $[A]$ means that $A$ is continuous at the 
hypersurface; a nonzero $[A]$ means that $A$ is discontinuous, and
$[A]$ gives the jump of $A$ across $\Sigma$.     

\subsection*{Intrinsic metric} 

It is easy to see that $\sigma_{AB}$ acts as a metric on $\Sigma$: 
A displacement on the hypersurface is given by $dx^\alpha_\pm =
k^\alpha_\pm\, d\lambda + \base{\alpha}{\pm A}\, d\theta^A$, and by
virtue of Eqs.~(\ref{2.3}) and (\ref{2.4}), we find 
\begin{equation}
ds^2_\Sigma = \sigma_{AB}\, d\theta^A d\theta^B 
\label{2.6}
\end{equation}
on either side of $\Sigma$. The condition (\ref{2.5}) therefore
ensures that the hypersurface possesses a well-defined intrinsic
geometry, a central requirement of the null-shell formalism. A
violation of Eq.~(\ref{2.5}) would indicate that the singularity of
the hypersurface is of a much stronger type than can be handled by
distributional techniques, and the physical interpretation of such a  
singular hypersurface would be much more delicate. 

It should be noted that the intrinsic metric of the null hypersurface
is two-dimensional. This is as it should be, as a displacement along a 
generator necessarily gives $ds^2 = 0$. This property, that the
hypersurface metric is not merely degenerate but explicitly
two-dimensional, is a definite advantage. For example, there is no 
difficulty in defining a metric inverse, which we denote
$\sigma^{AB}$, and which is the same on both sides of the
hypersurface. The inverse two-metric satisfies 
$\sigma^{AC} \sigma_{CB} = \delta^A_C$.       

\subsection*{Transverse vector} 

The tangent vectors $k^\alpha_\pm$ and $\base{\alpha}{\pm A}$ provide 
only a partial basis for the decomposition of vector fields on either
side of the hypersurface. We now complete the basis by adding a
transverse null vector $N^\alpha_\pm$ that satisfies 
\begin{equation}       
N_\alpha N^\alpha = 0, \qquad
N_\alpha k^\alpha = -1, \qquad
N_\alpha \base{\alpha}{A} = 0.  
\label{2.7}
\end{equation}
These four equations give a unique specification of the transverse
vector. The basis gives us the completeness relations   
\begin{equation}
g^{\alpha\beta} = -k^\alpha N^\beta - N^\alpha k^\beta 
+ \sigma^{AB} \base{\alpha}{A} \base{\beta}{B}  
\label{2.8} 
\end{equation}
for the inverse metric on either side of $\Sigma$ (in the coordinates  
$x_\pm^\alpha$); as was noted above, $\sigma^{AB}$ is the inverse of 
$\sigma_{AB}$. The completeness relations can be used to decompose a 
vector $A_\pm^\alpha$ in the specified basis: $A^\alpha =
g^{\alpha\beta} A_{\beta} = (-N^\beta A_\beta) k^\alpha + (-k^\beta
A_\beta) N^\alpha + (\sigma^{AB} \base{\beta}{B} A_\beta)
\base{\alpha}{A}$, where we have omitted the $\pm$ label for increased  
clarity.       

\subsection*{Congruence of timelike geodesics} 

The first key aspect of our reformulation of the null-shell
formalism was introduced previously: it is our choice of preferred
intrinsic coordinates which, as we shall see, provides the surface
stress-energy tensor with a simple description and a clear physical
interpretation. The second key aspect of our reformulation is the
introduction of an arbitrary congruence of timelike geodesics that
intersect the null hypersurface. As we shall see, the congruence will
give operational meaning to the distributional character of the
material stress-energy tensor.  

Each member of the congruence is thought of as the world line of
a freely-moving observer that intersects the surface layer and
performs measurements on it. The congruence then corresponds to a 
whole family of observers. We treat all such families equally and
do not attempt to introduce a preferred set of observers; the
congruence is therefore completely arbitrary. 

The timelike geodesics are parameterized by proper time $\tau$, which  
is adjusted so that $\tau = 0$ when a geodesic intersects $\Sigma$;
thus, $\tau < 0$ in ${\cal M}^-$ and $\tau > 0$ in ${\cal M}^+$. The 
vector field tangent to the congruence is $u^\alpha_\pm$ and is
normalized; a displacement along a single geodesic is given by
$dx^\alpha_\pm = u^\alpha_\pm\, d\tau$. To ensure that the congruence
is smooth at the hypersurface, we demand that $u^\alpha_\pm$ be ``the
same'' on both sides of $\Sigma$. This means that $u_\alpha
\base{\alpha}{a}$, the tangential projections of the vector field,
must be equal when evaluated on either side of the hypersurface:   
\begin{equation} 
\bigl[ -u_\alpha k^\alpha \bigr] = 0 = 
\bigl[ u_\alpha \base{\alpha}{A} \bigr]. 
\label{2.9}
\end{equation}
If, for example, $u^\alpha_-$ is specified in ${\cal M}^-$, then the
three conditions of Eq.~(\ref{2.9}) are sufficient (together with the 
geodesic equation) to determine the three independent components of
$u^\alpha_+$ in ${\cal M}^+$. We note that $-u_\alpha N^\alpha$, the
transverse projection of the vector field, is allowed to be
discontinuous across $\Sigma$.          

The proper-time parameter on the timelike geodesics can be viewed as  
as a scalar field $\tau(x^\alpha_\pm)$ defined in a normal convex
neighbourhood of $\Sigma$: Select a point $x^\alpha_\pm$ off the
hypersurface and locate the unique member of the congruence that
links this point to $\Sigma$; the value of the scalar field at
$x^\alpha_\pm$ is equal to the proper-time parameter of this geodesic
at that point. The hypersurface $\Sigma$ can then be described by the 
statement $\tau(x^\alpha_\pm) = 0$, and its normal vector
$k_\alpha^\pm$ will be proportional to the gradient of
$\tau(x^\alpha_\pm)$ evaluated on $\Sigma$. It is easy to check that
the appropriate expression is 
\begin{equation}
k_\alpha = - \bigl(-k_\mu u^\mu \bigr)\, 
\frac{\partial \tau}{\partial x^\alpha} \biggr|_{\Sigma} . 
\label{2.10} 
\end{equation}
We recall that the factor $-k_\mu u^\mu$ in Eq.~(\ref{2.10}) is
continuous at $\Sigma$.     

\subsection*{Transverse curvature} 

We have seen that the two-dimensional induced metric $\sigma_{AB}$ is
necessarily the same on both sides of the hypersurface. In the
presence of a surface layer, however, we can expect that the metric's
transverse derivative will be discontinuous across $\Sigma$. In
Sec.~\ref{III} we show that information about the metric's transverse
derivative is properly encoded in the {\it transverse curvature}
$C^\pm_{ab}$ defined by       
\begin{equation}
C_{ab} = -N_\alpha\, e^\alpha_{a;\beta} \base{\beta}{b} = C_{ba}.   
\label{2.11} 
\end{equation} 
In the presence of a surface layer, this object (which transforms as a 
scalar under transformations of the coordinates $x^\alpha_\pm$) is
discontinuous across $\Sigma$: $[C_{ab}] \neq 0$. The symmetry of
$C^\pm_{ab}$ follows from the identity $e^\alpha_{a;\beta}
\base{\beta}{b} = e^\alpha_{b;\beta} \base{\beta}{a}$, which states
that each basis vector $\base{\alpha}{a}$ is Lie transported along any
other basis vector. Note that $C_{ab}$ is the same object that was
denoted ${\cal K}_{ab}$ in the original work by Barrab\`es and Israel
\cite{BI}.   

\subsection*{Surface stress-energy tensor} 

The jump in the transverse curvature is directly related to the
stress-energy tensor of the surface layer. To display this
relationship we introduce the surface quantities 
\begin{eqnarray} 
\mu &=& -\frac{1}{8\pi}\, \sigma^{AB} \bigl[C_{AB}\bigr], 
\label{2.12} \\
j^A &=& \frac{1}{8\pi}\, \sigma^{AB} \bigl[ C_{\lambda B} \bigr],
\label{2.13} \\  
p &=& -\frac{1}{8\pi}\, \bigl[ C_{\lambda\lambda} \bigr]. 
\label{2.14}
\end{eqnarray}
We prove in Sec.~\ref{III} that the layer's surface stress-energy
tensor is given on either side of the hypersurface by   
\begin{equation}
S^{\alpha\beta} = \mu k^\alpha k^\beta + j^A \bigl
( k^\alpha \base{\beta}{A} + \base{\alpha}{A} k^\beta \bigr) 
+ p\, \sigma^{AB} \base{\alpha}{A} \base{\beta}{B}. 
\label{2.15}
\end{equation}
In this expression, the first term represents a flow of matter along
the null generators of the hypersurface, and the surface quantity
$\mu$ therefore represents a mass density. The second term represents
a flow of matter in the directions transverse to the generators, and 
the surface quantities $j^A$ therefore represent a current
density. Finally, the surface quantity $p$ clearly represents an
isotropic pressure. The surface layer can therefore be characterized
by four quantities: a density $\mu$, a current $j^A$, and an
isotropic pressure $p$. This provides the surface stress-energy tensor
with a transparent physical interpretation. As was noted before, this
clarity of interpretation comes from employing the null generators
to define a preferred flow on the hypersurface, to which the flow of
matter can be compared.      

We show in Sec.~\ref{III} that the complete stress-energy tensor
of the surface layer is given by     
\begin{equation}    
T_{\Sigma}^{\alpha\beta} = \bigl(-k_\mu u^\mu \bigr)^{-1} 
S^{\alpha\beta}\, \delta(\tau).   
\label{2.16}
\end{equation} 
In this expression, the factor $(-k_\mu u^\mu)^{-1}$ is continuous 
at $\Sigma$, and (as was discussed previously) the vector field
$u_\pm^\alpha$ is tangent to an arbitrary congruence of timelike
geodesics parameterized by $\tau$. The singular character of the
stress-energy tensor is revealed by the Dirac $\delta$-function, and
the congruence of timelike geodesics was introduced for the specific 
purpose of making the argument of the $\delta$-function
meaningful. Indeed, the fact that it is $\tau$ --- proper time as
measured by a family of freely-moving observers intersecting the
hypersurface --- and nothing else confers immediate operational
meaning to the prefactor $(-k_\mu u^\mu)^{-1}
S^{\alpha\beta}_\pm$. This choice of argument therefore clarifies 
the physical meaning of the surface stress-energy tensor.       

It is important to note that the presence of the factor  
$(-k_\mu u^\mu)^{-1}$ in front of $S^{\alpha\beta}_\pm$ in
Eq.~(\ref{2.16}) implies that $\mu$, $j^A$, and $p$ are not truly the
surface quantities that would be measured by the freely-moving
observers. The physically-measured surface quantities are instead 
given by  
\begin{eqnarray} 
\mu_{\rm pm} &=& \bigl(-k_\mu u^\mu \bigr)^{-1} \mu,
\label{2.17} \\ 
j^A_{\rm pm} &=& \bigl(-k_\mu u^\mu \bigr)^{-1} j^A, 
\label{2.18} \\ 
p_{\rm pm} &=& \bigl(-k_\mu u^\mu \bigr)^{-1} p.
\label{2.19}
\end{eqnarray}  
It is remarkable that the ambiguity associated with the arbitrary
choice of congruence is limited to a single multiplicative factor. The   
``bare'' quantities $\mu$, $j^A$, and $p$ are independent of this
choice, and it is often more convenient to work in terms of those.    

\subsection*{User manual} 

Our reformulation of the Barrab\`es-Israel null-shell formalism is now
complete. The formalism can be used as follows. First, locate the null
hypersurface $\Sigma$ in ${\cal M}^\pm$ and construct the intrinsic
coordinates $y^a = (\lambda,\theta^A)$; this step is the only
nonalgorithmic operation involved in the recipe. Second, construct the
tangent vectors $k^\alpha_\pm$, $\base{\alpha}{\pm A}$ and complete
the basis by computing the transverse null vector
$N^\pm_\alpha$. Third, compute the intrinsic metric $\sigma_{AB}$ and
verify that it is the same on both sides of the hypersurface. Fourth,
compute the transverse curvature $C^\pm_{ab}$ and its jump across
$\Sigma$. Fifth, and finally, compute the surface quantities $\mu$,
$j^A$, and $p$.        

\section{Derivations} 
\label{III} 

In this section we provide derivations of the main results presented
in Sec.~\ref{II}, namely Eqs.~(\ref{2.12})--(\ref{2.16}). Following
the Appendix of the Barrab\`es-Israel paper \cite{BI}, we introduce a
suitable coordinate system $x^\mu$ in a coordinate neighbourhood that
includes the null hypersurface $\Sigma$ and extends into both regions
${\cal M}^\pm$. These coordinates, which are distinct from the systems  
$x^\alpha_\pm$ encountered before, will be used for the derivations,
but as we saw in Sec.~\ref{II}, the final formulation of the
null-shell formalism will not refer to them.      

\subsection*{Singular part of the Riemann tensor} 

In the domain of our suitable coordinate system we write the metric as
a distribution-valued tensor:    
\begin{equation} 
g_{\alpha\beta} = g^+_{\alpha\beta} \Theta(\tau)
+ g^-_{\alpha\beta} \Theta(-\tau), 
\label{3.1}
\end{equation} 
where $\Theta(\tau)$ is the Heaviside step function and 
$g^\pm_{\alpha\beta}(x^\mu)$ is the metric in ${\cal M}^\pm$. To
ensure that the metric is well defined at $\tau = 0$ we assume that
in the coordinates $x^\mu$, $[g_{\alpha\beta}] = 0$; Eq.~(\ref{2.5})
is compatible with this requirement. We also impose $[k^\alpha] =
[\base{\alpha}{A}] = [N^\alpha] = [u^\alpha] = 0$, and these
continuity statements provide a precise meaning to the phrase 
``suitable coordinate system.''     

The metric of Eq.~(\ref{3.1}) generally gives rise to a singular
Riemann tensor, even when continuity of $g^\pm_{\alpha\beta}$ at
$\tau=0$ is enforced; the reason is that the {\it transverse
derivative} of the metric might be discontinuous at the
hypersurface. To compute the Riemann tensor we use Eq.~(\ref{2.10}) to
relate the gradient of $\tau$ to the null vector $k^\alpha$. The
result is 
\begin{equation}
R^\alpha_{\ \beta\gamma\delta} = 
R^{\ \ \alpha}_{+\ \beta\gamma\delta} \Theta(\tau) + 
R^{\ \ \alpha}_{-\ \beta\gamma\delta} \Theta(-\tau) + 
R_{\Sigma\ \beta\gamma\delta}^{\ \ \alpha}, 
\label{3.2}
\end{equation} 
where $R^{\ \ \alpha}_{\pm\ \beta\gamma\delta}$ denotes the Riemann
tensor in ${\cal M}^\pm$ constructed from $g^\pm_{\alpha\beta}$, and  
\begin{equation}
R_{\Sigma\ \beta\gamma\delta}^{\ \ \alpha} = -  
\bigl(-k_\mu u^\mu \bigr)^{-1}  
\bigl( \bigl[\Gamma^\alpha_{\ \beta\delta} \bigr] k_\gamma 
- \bigl[\Gamma^\alpha_{\ \beta\gamma} \bigr] k_\delta \bigr) \delta(\tau), 
\label{3.3}
\end{equation}
is the singular part of the Riemann tensor. Here, 
$[\Gamma^\alpha_{\ \beta\gamma}]$ denotes the jump in the affine
connection across $\Sigma$.    

We must now characterize the discontinuous behaviour of 
$g_{\alpha\beta,\gamma}$. The condition $[g_{\alpha\beta}]=0$
guarantees that the tangential derivatives of the metric are
continuous: $[g_{\alpha\beta,\gamma}] k^\gamma = 0 =
[g_{\alpha\beta,\gamma}]\, \base{\gamma}{C}$. The only possible
discontinuity is therefore in $g_{\alpha\beta,\gamma} N^\gamma$, the
transverse derivative of the metric. In view of Eq.~(\ref{2.7}) 
we conclude that there exists a tensor field
$\gamma_{\alpha\beta}$ such that   
\begin{equation}
\bigl[ g_{\alpha\beta,\gamma} \bigr] = -\gamma_{\alpha\beta} k_\gamma.   
\label{3.4} 
\end{equation}
This tensor is given explicitly by $\gamma_{\alpha\beta} = 
[ g_{\alpha\beta,\gamma} ] N^\gamma$ and it follows that    
\begin{equation} 
\bigl[\Gamma^\alpha_{\ \beta\gamma} \bigr] = -\frac{1}{2} 
\bigl( \gamma^\alpha_{\ \beta} k_\gamma 
+ \gamma^\alpha_{\ \gamma} k_\beta
- \gamma_{\beta\gamma} k^\alpha \bigr).  
\label{3.5}
\end{equation}
Substituting this into Eq.~(\ref{3.3}) gives 
\begin{eqnarray}
R_{\Sigma\ \beta\gamma\delta}^{\ \ \alpha} &=& \frac{1}{2}   
\bigl(-k_\mu u^\mu \bigr)^{-1} \bigl( 
\gamma^\alpha_{\ \delta} k_\beta k_\gamma 
- \gamma_{\beta\delta} k^\alpha k_\gamma 
\nonumber \\ & & \mbox{} 
- \gamma^\alpha_{\ \gamma} k_\beta k_\delta  
+ \gamma_{\beta\gamma} k^\alpha k_\delta \bigr) \delta(\tau), 
\label{3.6}
\end{eqnarray} 
and we see that $k^\alpha$ and $\gamma_{\alpha\beta}$ give a complete   
characterization of the singular part of the Riemann tensor. 

\subsection*{Surface stress-energy tensor} 

From Eq.~(\ref{3.6}) we form the singular part of the Einstein tensor,
and the Einstein field equations give us the singular part of the
stress-energy tensor:   
\begin{equation}
T_\Sigma^{\alpha\beta} = \bigl(-k_\mu u^\mu \bigr)^{-1}
S^{\alpha\beta}\, \delta(\tau), 
\label{3.7}
\end{equation}
where 
\begin{equation} 
S^{\alpha\beta} = \frac{1}{16\pi} \bigl( 
k^\alpha \gamma^\beta_{\ \mu} k^\mu
+ k^\beta \gamma^\alpha_{\ \mu} k^\mu
- \gamma^\mu_{\ \mu} k^\alpha k^\beta 
- \gamma_{\mu\nu} k^\mu k^\nu g^{\alpha\beta} \bigr)  
\label{3.8}
\end{equation} 
is the surface stress-energy tensor of the null shell, up to a 
factor $(-k_\mu u^\mu)$ that depends on the choice of observers making  
measurements on the shell. 

The expression for the surface stress-energy tensor can be simplified
if we decompose it into the basis $(k^\alpha, N^\alpha,
\base{\alpha}{A})$. For this purpose we introduce the projections     
\begin{equation}  
\gamma_A \equiv \gamma_{\alpha\beta}\, \base{\alpha}{A} k^\beta,
\qquad  
\gamma_{AB} \equiv \gamma_{\alpha\beta}\, \base{\alpha}{A}
\base{\beta}{B},  
\label{3.9} 
\end{equation} 
and we use the completeness relations of Eq.~(\ref{2.8}) to find that
the vector $\gamma^\alpha_{\ \mu} k^\mu$ admits the decomposition  
\begin{eqnarray} 
\gamma^\alpha_{\ \mu} k^\mu &=& \frac{1}{2} \bigl( \gamma^\mu_{\ \mu} -
\sigma^{AB} \gamma_{AB} \bigr) k^\alpha 
- \bigl(\gamma_{\mu\nu} k^\mu k^\nu \bigr) N^\alpha 
\nonumber \\ & & \mbox{} 
+ \bigl( \sigma^{AB} \gamma_B \bigr)\, \base{\alpha}{A}.   
\label{3.10}
\end{eqnarray} 
Substituting this into Eq.~(\ref{3.8}) and involving once more the
completeness relations, we arrive at our final expression for the
surface stress-energy tensor:  
\begin{equation}
S^{\alpha\beta} = \mu k^\alpha k^\beta + j^A \bigl( k^\alpha
\base{\beta}{A} + \base{\alpha}{A} k^\beta \bigr) + p\, \sigma^{AB} 
\base{\alpha}{A} \base{\beta}{B},  
\label{3.11} 
\end{equation}
where
\begin{equation} 
\mu \equiv -\frac{1}{16\pi}\, \bigl( \sigma^{AB} \gamma_{AB} \bigr) 
\label{3.12}
\end{equation} 
is the surface density,  
\begin{equation} 
j^A \equiv \frac{1}{16\pi}\, \bigl( \sigma^{AB} \gamma_{B} \bigr) 
\label{3.13}
\end{equation}
is the surface current, and 
\begin{equation} 
p \equiv -\frac{1}{16\pi}\, 
\bigl( \gamma_{\alpha\beta} k^\alpha k^\beta \bigr) 
\label{3.14}
\end{equation}
is the isotropic surface pressure.  

The surface stress-energy tensor of Eq.~(\ref{3.11}) is expressed  
in the suitable coordinates $x^\mu$. This is, however, a tensorial 
equation involving vectors ($k^\alpha$ and $\base{\alpha}{A}$) and
scalars ($\mu$, $j^A$, and $p$). This equation can therefore be
expressed in any coordinate system. In particular, when viewed from
${\cal M}^\pm$, the surface stress-energy tensor can be expressed in
the original coordinates $x^\alpha_\pm$. This means that
Eq.~(\ref{3.7}) is formally identical to Eq.~(\ref{2.16}), while 
Eq.~(\ref{3.11}) is identical to Eq.~(\ref{2.15}).   

\subsection*{Intrinsic formulation} 

The surface quantities of Eqs.~(\ref{3.12})--(\ref{3.14}) are
constructed from the tensor $\gamma_{\alpha\beta}$ which is defined  
only in the suitable coordinates $x^\mu$. As it might be
grossly impractical to construct such a coordinate system, it is
important to find an independent way of obtaining these quantities. It
is at this stage that we introduce the transverse curvature $C_{ab}$ 
defined by Eq.~(\ref{2.11}). Recalling from Eq.~(\ref{2.7}) that
$N_\alpha \base{\alpha}{a}$ are constants (either zero or minus one), 
we express the transverse curvature as 
\begin{equation} 
C_{ab} = N_{\alpha;\beta} \base{\alpha}{a} \base{\beta}{b}
\label{3.15}
\end{equation} 
and we seek to relate its jump to the surface quantities. In the next
paragraph we will use the suitable coordinates $x^\mu$ to establish
the general validity of the scalar relations
(\ref{2.12})--(\ref{2.14}). This will bring our derivations to a 
close.    

In the suitable coordinates, the jump in the transverse curvature is 
given by  
\begin{eqnarray}
\bigl[ C_{ab} \bigr] &=& \bigl[ N_{\alpha;\beta} \bigr]\, 
\base{\alpha}{a} \base{\alpha}{b} 
= - \bigl[ \Gamma^\gamma_{\ \alpha\beta} \bigr] N_\gamma   
\base{\alpha}{a} \base{\alpha}{b} \nonumber \\ 
&=& \frac{1}{2}\, \gamma_{\alpha\beta}\, 
\base{\alpha}{a} \base{\alpha}{b}, 
\label{3.16} 
\end{eqnarray}
where we have used Eq.~(\ref{3.5}) and the fact that $k^\alpha$ is 
orthogonal to $\base{\alpha}{a}$. We therefore have
$[C_{\lambda\lambda}] = \frac{1}{2} \gamma_{\alpha\beta} k^\alpha
k^\beta$, $[C_{A \lambda}] = \frac{1}{2} \gamma_{\alpha\beta}
\base{\alpha}{A} k^\beta \equiv \frac{1}{2} \gamma_A$, and $[C_{AB}] = 
\frac{1}{2} \gamma_{\alpha\beta}\base{\alpha}{A}\base{\beta}{B} \equiv 
\frac{1}{2} \gamma_{AB}$, where we have involved
Eq.~(\ref{3.9}). Finally, using Eqs.~(\ref{3.12})--(\ref{3.14}) we
find that the surface quantities can indeed be expressed as in 
Eqs.~(\ref{2.12})--(\ref{2.14}).  

We have established that the shell's surface quantities can all be
computed directly in terms of the induced metric $\sigma_{AB}$ and the
jump of the transverse curvature $C_{ab}$ across $\Sigma$. This
confirms that the suitable coordinates $x^\mu$ are not needed in
practical applications of the null-shell formalism.   

\section{Parameterization of the null generators} 
\label{IV} 

In this section we discuss two technical issues related to the
parameterization of the null generators of the singular
hypersurface. In the first subsection we justify a statement made
earlier, that in general $\lambda$ cannot be an affine parameter on
both sides of the hypersurface; our presentation here is based
directly on Sec.~III of the Barrab\`es-Israel paper \cite{BI}, but for 
completeness we adapt their discussion to the notation of this 
paper. In the second subsection we examine the transformation 
properties of the surface quantities under a reparameterization of the
null generators.  

\subsection*{Affine parameterization of the null generators}   

Whether or not $\lambda$ is an affine parameter can be decided by
computing $\kappa_\pm$, the ``acceleration''  of the null  
vector $k_\pm^\alpha$. This is defined on either side of the
hypersurface by 
\begin{equation}
k^\alpha_{\ ;\beta} k^\beta = \kappa k^\alpha, 
\label{4.1}
\end{equation} 
and $\lambda$ will be an affine parameter on the ${\cal M}^\pm$ 
side of $\Sigma$ if $\kappa_\pm = 0$. According to Eq.~(\ref{2.7}),    
$\kappa = - N_\alpha k^{\alpha}_{\ ;\beta} k^\beta = -N_\alpha
e^{\alpha}_{\lambda;\beta} \base{\beta}{\lambda} =
C_{\lambda\lambda}$, where we have also used Eqs.~(\ref{2.2}) and 
(\ref{2.11}). Equation (\ref{2.14}) then relates the discontinuity in
the acceleration to the surface pressure: 
\begin{equation} 
[ \kappa ] = - 8\pi\, p. 
\label{4.2} 
\end{equation} 
We conclude that $\lambda$ can be an affine parameter on both sides of
$\Sigma$ only when the surface layer has a vanishing surface
pressure. When $p \neq 0$, on the other hand, $\lambda$ can be chosen
to be an affine parameter on {\it one} side of the hypersurface, but
it will not be an affine parameter on the other side. 

Additional insight into this matter can be gained from Raychaudhuri's 
equation, 
\begin{equation} 
\frac{d\theta}{d\lambda} + \frac{1}{2}\theta^2 + \sigma^{\alpha\beta}
\sigma_{\alpha\beta} = \kappa\, \theta  
- 8\pi T_{\alpha\beta} k^\alpha k^\beta, 
\label{4.3}
\end{equation} 
which describes the evolution of the congruence of null generators
\cite{wald, carter}; the equation holds on either side of the
hypersurface. Here, $\theta$ and $\sigma_{\alpha\beta}$ are the
expansion and shear of the congruence, respectively, and we have set
the vorticity to zero because the congruence is hypersurface
orthogonal. Because it depends only on the intrinsic geometry of the
hypersurface, the left-hand side of Raychaudhuri's equation is
necessarily continuous across the shell. Continuity of the
right-hand side therefore implies   
\begin{equation}
[ \kappa ] \theta = 8\pi 
\bigl[ T_{\alpha\beta} k^\alpha k^\beta \bigr].
\label{4.4} 
\end{equation}
This relation shows that $[\kappa] \neq 0$ (and therefore $p \neq 0$)
whenever the component $T_{\alpha\beta} k^\alpha k^\beta$ of the
stress-energy tensor is discontinuous across the shell. We conclude
that $\lambda$ cannot be an affine parameter on both sides of $\Sigma$
when $[T_{\alpha\beta} k^\alpha k^\beta] \neq 0$. (Notice that this 
conclusion breaks down when $\theta = 0$, that is, when the shell is 
stationary.)

Recalling that the expansion $\theta$ is equal to the fractional rate
of change of the congruence's cross-sectional area, we find that with
the help of Eq.~(\ref{4.2}), Eq.~(\ref{4.4}) can be expressed as
$-p\theta = [T_{\alpha\beta} k^\alpha k^\beta]$, or  
\begin{equation}
- p\, \frac{d}{d\lambda}{dS} 
= \bigl[ T_{\alpha\beta} k^\alpha k^\beta \bigr] dS,  
\label{4.5} 
\end{equation} 
where $dS$ is an element of area transverse to the generators. This
equation has a straightforward interpretation: The left-hand side 
represents the work done on the shell as it expands or contracts,
while the right-hand side is the energy absorbed by the shell from 
its surroundings; Eq.~(\ref{4.5}) therefore states that all of the
absorbed energy goes into work. 

\subsection*{Reparameterization of the null generators}    

We now investigate how a change of parameterization affects the
surface density $\mu$, surface current $j^A$, and surface pressure $p$
of the null shell. Because each generator can be reparameterized
independently of any other generator, we must consider transformations
of the form   
\begin{equation}  
\lambda \to \bar{\lambda}(\lambda,\theta^A). 
\label{4.6}
\end{equation} 
The question before us is: How do the surface quantities change under
such a transformation?  

To answer this we need to work out how the transformation of
Eq.~(\ref{4.6}) affects the vectors $k^\alpha$, $\base{\alpha}{A}$,
and $N^\alpha$. We first note that the differential form of
Eq.~(\ref{4.6}) is    
\begin{equation}
d\bar{\lambda} = e^\beta\, d\lambda + c_A\, d\theta^A, 
\label{4.7}
\end{equation}
where
\begin{equation}
e^\beta \equiv \biggl( \frac{\partial \bar{\lambda}}{\partial \lambda}  
\biggr)_{\!\! \theta^A}, \qquad
c_A \equiv \biggl( \frac{\partial \bar{\lambda}}{\partial \theta^A}   
\biggr)_{\!\! \lambda};  
\label{4.8}
\end{equation}  
both $e^\beta$ and $c_A$ depend on $y^a = (\lambda,\theta^A)$, but
because they depend on the intrinsic coordinates only, we have that 
$[e^\beta] = 0 = [c_A]$. A displacement within the hypersurface can
then be described either by
\begin{equation}
dx^\alpha = k^\alpha\, d\lambda + \base{\alpha}{A}\, d\theta^A, 
\label{4.9}
\end{equation} 
where $k^\alpha = (\partial x^\alpha/\partial\lambda)_{\theta_A}$ and 
$\base{\alpha}{A} = (\partial x^\alpha/\partial\theta^A)_\lambda$, or
by   
\begin{equation}
dx^\alpha = \bar{k}^{\alpha}\, d\bar{\lambda} 
+ \bar{e}^{\alpha}_{A}\, d\theta^A,  
\label{4.10} 
\end{equation} 
where $\bar{k}^{\alpha} = 
(\partial x^\alpha/\partial\bar{\lambda})_{\theta_A}$ 
and $\bar{e}^{\alpha}_{A} = 
(\partial x^\alpha/\partial\theta^A)_{\bar{\lambda}}$; these relations  
hold on either side of $\Sigma$, in the relevant coordinate system 
$x^\alpha_\pm$. Using Eq.~(\ref{4.7}), it is easy to see that the
tangent vectors transform as   
\begin{equation}
\bar{k}^{\alpha} = e^{-\beta}\, k^\alpha, \qquad
\bar{e}^{\alpha}_{A} = \base{\alpha}{A}  
- c_A e^{-\beta}\, k^\alpha  
\label{4.11}
\end{equation}
under the reparameterization of Eq.~(\ref{4.6}). It may be 
checked that the new basis vectors satisfy the orthogonality relations
of Eq.~(\ref{2.3}), and that the induced metric is
invariant under this transformation: $\bar{\sigma}_{AB} \equiv
g_{\alpha\beta} \bar{e}^{\alpha}_{A} \bar{e}^{\beta}_{B} =
g_{\alpha\beta} \base{\alpha}{A} \base{\beta}{B} \equiv
\sigma_{AB}$. To preserve the relations (\ref{2.7}) we let the new
transverse null vector be  
\begin{equation}  
\bar{N}^{\alpha} = e^\beta\, N^\alpha + \frac{1}{2} c^A c_A
e^{-\beta}\, k^\alpha - c^A \base{\alpha}{A}, 
\label{4.12} 
\end{equation} 
where $c^A = \sigma^{AB} c_B$. This ensures that the completeness
relations of Eq.~(\ref{2.8}) take the same form in the new basis.  

It is a straightforward (but slightly tedious) task to compute how 
the transverse curvature $C_{ab}$ changes under a reparameterization
of the generators, and to then compute how the surface quantities 
transform. We find 
\begin{eqnarray}
\bar{\mu} &=& e^\beta \mu + 2 c_A j^A 
+ c^A c_A e^{-\beta} p, 
\label{4.13} \\
\bar{\jmath}^A &=& j^A + c^A e^{-\beta} p,    
\label{4.14} \\ 
\bar{p} &=& e^{-\beta} p. 
\label{4.15}  
\end{eqnarray} 
These transformations, together with Eq.~(\ref{4.11}), imply that   
the surface stress-energy tensor becomes $\bar{S}^{\alpha\beta} =
e^{-\beta} S^{\alpha\beta}$. We also have $(-\bar{k}_\mu u^\mu)^{-1}  
= e^\beta (-k_\mu u^\mu \bigr)$, and these results imply that the  
combination $(-k_\mu u^\mu)^{-1} S^{\alpha\beta}$ is {\it invariant}
under the reparameterization. This, finally, establishes the
invariance of $T_\Sigma^{\alpha\beta}$.   

As a final remark, we note that under the reparameterization of
Eq.~(\ref{4.6}), the physically-measured surface quantities of
Eqs.~(\ref{2.17})--(\ref{2.19}) transform as 
\begin{eqnarray}
\bar{\mu}_{\rm pm} &=& e^{2\beta} \mu_{\rm pm} 
+ 2 c_A e^\beta j^A_{\rm pm} + c^A c_A p_{\rm pm}, 
\label{4.16} \\
\bar{\jmath}^A_{\rm pm} &=& e^\beta j^A_{\rm pm} + c^A p_{\rm pm},   
\label{4.17} \\  
\bar{p}_{\rm pm} &=& p_{\rm pm}; 
\label{4.18} 
\end{eqnarray} 
we see in particular that the physically-measured surface pressure is
an invariant.    

\section{Applications} 
\label{V} 

In this section we consider four simple applications of the
null-shell formalism. Those are designed to illustrate the basic
methods involved and highlight some aspects of the physics of null
surface layers.   

\subsection*{Imploding spherical shell} 

For our first application of the null-shell formalism, we look at the
gravitational collapse of a thin spherical shell. We imagine that the
collapse proceeds at the speed of light, so that the world surface of
the shell coincides with a null hypersurface $\Sigma$. We take
spacetime to be flat inside the shell (in ${\cal M}^-$), and write the
metric there as   
\begin{equation}
ds^2_- = -dt_-^2 + dr^2 + r^2\, d\Omega^2, 
\label{5.1}
\end{equation} 
in terms of spatial coordinates $(r,\theta,\phi)$ and a time 
coordinate $t_-$. The metric outside the shell (in ${\cal M}^+$) is  
the Schwarzschild solution, 
\begin{equation} 
ds_+^2 = -f\, dt_+^2 + f^{-1}\, dr^2 + r^2\, d\Omega^2, 
\label{5.2}
\end{equation} 
expressed in the same spatial coordinates but in terms of a distinct
time $t_+$; we have $f = 1-2M/r$ and $M$ designates the gravitational 
mass of the imploding shell.    

As seen from ${\cal M}^-$, the null hypersurface is described by the
equation $t_- + r \equiv v_- = \mbox{constant}$, which implies that
the induced metric on $\Sigma$ is given by $ds^2_\Sigma = r^2\,  
d\Omega^2$. As see from ${\cal M}^+$, on the other hand, the 
hypersurface is described by $t_+ + r^*(r) \equiv v_+ =
\mbox{constant}$, where $r^*(r) = \int f^{-1}\, dr = r + 2M\ln(r/2M -
1)$, and this gives rise to the same induced metric. From these  
considerations we conclude that it was permissible to express the
metrics of ${\cal M}^\pm$ in terms of the same spatial coordinates
$(r,\theta,\phi)$, but that $t_+$ cannot be equal to $t_-$. The
shell's intrinsic metric is   
\begin{equation} 
\sigma_{AB}\, d\theta^A d\theta^B = \lambda^2 \bigl( d\theta^2 +
\sin^2\theta\, d\phi^2 \bigr), 
\label{5.3}
\end{equation} 
where we have set $\theta^A = (\theta,\phi)$ and identified $-r$ with 
the parameter $\lambda$ on the null generators of the hypersurface. We 
shall see that here, $\lambda$ is an affine parameter on both sides of
$\Sigma$.    

As seen from ${\cal M}^-$, the parametric equations $x_-^\alpha =   
x_-^\alpha(\lambda,\theta^A)$ that describe the hypersurface have the 
explicit form $t_- = v_- + \lambda$, $r = -\lambda$, $\theta = 
\theta$, and $\phi = \phi$. These give us the tangent vectors 
$k^\alpha \partial_\alpha = \partial_t - \partial_r$,
$\base{\alpha}{\theta} \partial_\alpha = \partial_\theta$, and
$\base{\alpha}{\phi} \partial_\alpha = \partial_\phi$, and the basis
is completed by the transverse vector $N_\alpha\, dx^\alpha =
-\frac{1}{2}( dt - dr )$. From all this and Eq.~(\ref{2.11}) we find
that the nonvanishing components of the transverse curvature are   
\begin{equation}
C^-_{AB} = \frac{1}{2r}\, \sigma_{AB}. 
\label{5.4}
\end{equation} 
The fact that $C^-_{\lambda\lambda} = 0$ confirms that $\lambda \equiv 
-r$ is an affine parameter on the ${\cal M}^-$ side of $\Sigma$.       

As seen from ${\cal M}^+$, the parametric equations are $t_+ = v_+ -
r^*(-\lambda)$, $r = -\lambda$, $\theta = \theta$, and 
$\phi = \phi$. The basis vectors are   
$k^\alpha \partial_\alpha = f^{-1} \partial_t - \partial_r$, 
$\base{\alpha}{\theta} \partial_\alpha = \partial_\theta$,
$\base{\alpha}{\phi} \partial_\alpha = \partial_\phi$, and
$N_\alpha\, dx^\alpha = -\frac{1}{2}( f\, dt - dr)$. The nonvanishing
components of the transverse curvature are now    
\begin{equation}
C^+_{AB} = \frac{f}{2r}\, \sigma_{AB}.  
\label{5.5}
\end{equation} 
The fact that $C^+_{\lambda\lambda} = 0$ confirms that $\lambda \equiv  
-r$ is an affine parameter on the ${\cal M}^+$ side of $\Sigma$;   
$\lambda$ is therefore an affine parameter on both sides.  

The angular components of the transverse curvature are discontinuous 
across the shell: $[C_{AB}] = -(M/r^2)\sigma_{AB}$. According to
Eqs.~(\ref{2.12})--(\ref{2.14}), this means that the shell has a
vanishing surface current $j^A$ and a vanishing surface pressure $p$,
but that its surface density is 
\begin{equation}
\mu = \frac{M}{4\pi r^2}. 
\label{5.6}
\end{equation} 
We have therefore obtained the very sensible result that the surface 
density of an imploding null shell is equal to its gravitational mass 
divided by its (ever decreasing) surface area. Notice that 
$\mu_{\rm pm} = \mu$ for observers at rest in ${\cal M}^-$. Because of
the gravitational action of the null shell, however, these observers
do not remain at rest after crossing over to the ${\cal M}^+$ side: A
simple calculation, based on Eq.~(\ref{2.9}), reveals that an observer
at rest before crossing the shell will move according to $dr/d\tau =
-(\gamma^2 -f)^{1/2}$ after crossing the shell; the constant $\gamma$
varies from observer to observer, and is related by $\gamma = 1 -
M/r_\Sigma$ to the radius $r_\Sigma$ at which a given observer crosses 
the hypersurface.   

\subsection*{Spherical null shells} 

In our second application of the null-shell formalism, we extend  
the results of the preceding subsection to a situation in which the 
surface layer is immersed in a general spherically-symmetric 
spacetime. Our results here are identical to those first presented by 
Barrab\`es and Israel in Sec.~IV of their paper \cite{BI}. We adapt
their discussion to the notation of this paper for completeness, as
spherically-symmetric situations are of extreme practical
importance. We assume that the surface layer is either expanding 
or contracting, and we exclude stationary hypersurfaces from our
considerations. For a discussion of the interesting special case of
horizon-straggling shells, we refer the reader to the original 
Barrab\`es-Israel work.      

The metric in ${\cal M}^\pm$ is expressed in the general form
\begin{equation}
ds^2 = -e^{\psi}\, dw \bigl( f e^{\psi}\, dw + 2 \zeta\, dr \bigr) +
r^2\, d\Omega^2, 
\label{5.7}
\end{equation}
where $\psi_\pm$ and $f_\pm$ are two arbitrary functions of the
coordinates $w_\pm$ and $r$; we also have the parameter $\zeta = 
\pm 1$ whose meaning will be explained shortly. (Notice that the 
same $r$ is used as a coordinate in both regions.) It is convenient to
introduce the mass functions $m_\pm(w_\pm,r)$ defined by the relation    
\begin{equation}
f = 1 - 2m/r.   
\label{5.8}
\end{equation}   
The hypersurfaces $w_\pm = \mbox{constant}$ are null. They are
expanding ($r$ increases along the generators) if $\zeta = +1$, and
$w_\pm$ (which normally would be denoted $u_\pm$) is then properly
interpreted as a retarded-time coordinate. On the other hand, the null
hypersurfaces are contracting ($r$ decreases along the generators) if
$\zeta = -1$, and $w_\pm$ (which normally would be denoted $v_\pm$) is
then properly interpreted as an advanced-time coordinate.  

The boundary between ${\cal M}^-$ and ${\cal M}^+$ is taken to be a
particular hypersurface $w_\pm = \mbox{constant}$. According to
Eq.~(\ref{5.7}), its intrinsic metric is given by 
\begin{equation} 
\sigma_{AB}\, d\theta^A d\theta^B = r^2 (d\theta^2 + \sin^2\theta\,
d\phi^2), 
\label{5.9}
\end{equation}
which is the same on both sides of $\Sigma$. We have identified
$\theta^A$ with the two angles $(\theta, \phi)$, and we take $\lambda
\equiv \zeta r$ to be the parameter on the null generators. Notice
that $\lambda$ always increases along the generators. 

The parametric equations that describe the hypersurface are $w_\pm = 
\mbox{constant}$, $r = \zeta \lambda$, $\theta = \theta$, and $\phi =
\phi$. The tangent vectors are then 
\begin{equation} 
k^\alpha \partial_\alpha = \zeta\, \partial_r, \qquad
\base{\alpha}{\theta} \partial_\alpha = \partial_\theta, \qquad   
\base{\alpha}{\phi} \partial_\alpha = \partial_\phi
\label{5.10}
\end{equation}
on either side of the hypersurface. The basis is completed by the
transverse null vector 
\begin{equation}
N_\alpha\, dx^\alpha = -\frac{1}{2} \bigl( f e^{\psi}\, dw 
+ 2\zeta\, dr \bigr).  
\label{5.11}
\end{equation}
We may now use Eq.~(\ref{2.11}) to compute the transverse
curvature. The nonvanishing components are 
\begin{equation} 
C_{\lambda\lambda} \equiv \kappa 
= \zeta \frac{\partial \psi}{\partial r}, \qquad 
C_{AB} = -\zeta \frac{f}{2r}\, \sigma_{AB}. 
\label{5.12}
\end{equation}
We recall from Sec.~\ref{IV} that $\kappa_\pm$ is the ``acceleration''
of the null vector $k^\alpha_\pm$; from the first equation we
therefore learn that $\lambda$ is an affine parameter on the 
${\cal M}^\pm$ side of $\Sigma$ when 
$\partial \psi_\pm / \partial r$ = 0. 

According to Eqs.~(\ref{2.12})--(\ref{2.14}) and (\ref{5.12}), the
only nonvanishing surface quantities are $\mu$ and $p$. Using
Eq.~(\ref{5.8}) we find that the mass density is given by 
\begin{equation}
\mu = (-\zeta) \frac{ \bigl[ m \bigr] }{4\pi r^2}; 
\label{5.13}
\end{equation} 
this is the difference in the mass function across the shell,
$[m](r)$, divided by the shell's surface area, $4\pi r^2$. Assuming
that $\mu$ is positive, we see that $[m] < 0$ if the hypersurface is 
expanding --- the shell removes energy from the centre --- whereas
$[m] > 0$ if the hypersurface is contracting --- the shell then 
brings energy to the centre. The shell's surface pressure is given by  
\begin{equation} 
p = (-\zeta) \frac{1}{8\pi} \biggl[ \frac{\partial \psi}{\partial r}
\biggr]. 
\label{5.14}
\end{equation} 
This result can be related to the first law of thermodynamics if we
note that $[\partial \psi/\partial r] = 4\pi r [T_{\alpha\beta}
k^\alpha k^\beta]$ according to the Einstein field equations, and that
the expansion of the null generators is given by $\theta = 
k^\alpha_{\ ;\alpha} - \kappa = 2\xi/r$. Equation (\ref{5.13}) is then
seen to be equivalent to the statement $p\theta = - [T_{\alpha\beta}
k^\alpha k^\beta]$ previously displayed in Eq.~(\ref{4.5}).             

\subsection*{Accreting black hole} 

In our third application of the null-shell formalism, we consider a 
nonrotating black hole of mass $(M-m)$ which suddenly acquires
additional material of mass $m$ and angular momentum $J \equiv 
aM \ll M^2$. We describe the accreting material by a singular 
stress-energy tensor supported on a null hypersurface $\Sigma$. 

The spacetime in the future of $\Sigma$ --- in ${\cal M}^+$ --- is
that of a slowly rotating black hole of mass $M$ and (small) angular
momentum $aM$. We write the metric in ${\cal M}^+$ as 
\begin{equation} 
ds^2_+ = -f\, dt^2 + f^{-1}\, dr^2 + r^2\, d\Omega^2 
- \frac{4Ma}{r}\, \sin^2\theta\, dtd\phi,   
\label{5.15}
\end{equation} 
where $f = 1-2M/r$; this is the slow-rotation limit of the Kerr
metric, and throughout this subsection we will work consistently to
first order in the small parameter $a$. 

As seen from ${\cal M}^+$, the null hypersurface is described 
by $v \equiv t + r^* = 0$, where $r^* = \int f^{-1}\, dr = r 
+ 2M\ln(r/2M - 1)$. In the slow-rotation limit, every surface $v = 
\mbox{constant}$ is null, and it follows that the vector $k^\alpha =
g^{\alpha\beta}(-\partial_\beta v)$ is normal to $\Sigma$ and tangent
to its null generators. We have  
\begin{equation} 
k^\alpha \partial_\alpha = \frac{1}{f}\, \partial_t - \partial_r +
\frac{2Ma}{r^3 f}\, \partial_\phi, 
\label{5.16}
\end{equation} 
and from this expression we deduce four important properties of the   
generators. First, the generators are affinely parameterized by 
$\lambda \equiv -r$. Second, as measured by inertial observers at
infinity, the generators move with an (ever increasing) angular
velocity    
\begin{equation} 
\frac{d\phi}{dt} \equiv
\Omega_{\rm generators} =  \frac{2Ma}{r^3}. 
\label{5.17}
\end{equation}  
Third, $\theta$ is constant on each generator. And fourth, integration 
of $d\phi/(-dr) = 2Ma/(r^3 f)$ reveals that   
\begin{equation} 
\psi \equiv \phi + \frac{a}{r} 
\Bigl( 1 + \frac{r}{2M} \ln f \Bigr) 
\label{5.18}
\end{equation} 
also is constant on the generators. 

We shall use $y^a = (\lambda \equiv -r,\theta,\psi)$ as coordinates on
$\Sigma$, as these are well adapted to the generators. Remembering
that $dt = -dr/f$ and $d\phi = d\psi - (2Ma/r^3 f)\, dr$ on $\Sigma$,
we find that the induced metric is    
\begin{equation} 
\sigma_{AB}\, d\theta^A d\theta^B = r^2 \bigl( d\theta^2 +
\sin^2\theta\, d\psi^2 \bigr), 
\label{5.19}
\end{equation} 
and we see that the hypersurface is intrinsically spherical.  

The parametric description of $\Sigma$, as seen from ${\cal M}^+$, is 
$x^\alpha(-r,\theta,\psi)$, and from this we form the tangent vectors
$\base{\alpha}{\lambda} = k^\alpha$, $\base{\alpha}{\theta} =
\delta^\alpha_{\theta}$, and $\base{\alpha}{\psi} =
\delta^\alpha_{\phi}$. The basis is completed by the transverse null
vector $N_\alpha\, dx^\alpha = \frac{1}{2}(- f\, dt + dr)$. From  
Eq.~(\ref{2.11}) we obtain 
\begin{equation} 
C^+_{\lambda \psi} = \frac{3Ma}{r^2}\, \sin^2\theta, \qquad
C^+_{AB} = \frac{f}{2r}\, \sigma_{AB} 
\label{5.20}
\end{equation} 
for the nonvanishing components of the transverse curvature.   

The spacetime in the past of $\Sigma$ --- in ${\cal M}^-$ --- is that
of a nonrotating black hole of mass $(M-m)$. Here we write the metric
as  
\begin{equation} 
ds_-^2 = -F\, d\bar{t}^{\, 2} + F^{-1}\, dr^2  
+ r^2 \bigl( d\theta^2 + \sin^2\theta\, d\psi^2 \bigr), 
\label{5.21}
\end{equation} 
in terms of a distinct time coordinate $\bar{t}$ and the angles  
$\theta$ and $\psi$; we also have $F \equiv 1 - 2(M-m)/r$. This  
choice of angular coordinates reflects the fact that
zero-angular-momentum observers in ${\cal M}^-$ corotate with the
shell's null generators; this is a manifestation of the dragging of
inertial frames. As we shall see presently, this choice of coordinates
is dictated by continuity of the induced metric at $\Sigma$.      

The mathematical description of the hypersurface, as seen from 
${\cal M}^-$, is identical to its external description provided  
that we make the substitutions $t \to \bar{t}$, $\phi \to \psi$,  
$M \to M-m$, and $a \to 0$. According to this, the induced metric on   
$\Sigma$ is still given by $ds^2_\Sigma = r^2 (d\theta^2 +
\sin^2\theta\, d\psi^2)$, as required. The basis vectors are now
$k^\alpha \partial_\alpha = F^{-1}\partial_{\bar{t}} - \partial_r$,   
$\base{\alpha}{\theta} \partial_\alpha = \partial_\theta$, 
$\base{\alpha}{\psi} \partial_\alpha = \partial_\psi$, and 
$N_\alpha\, dx^\alpha = \frac{1}{2}(- F\, d\bar{t} + dr)$. This
gives us  
\begin{equation}
C^-_{AB} = \frac{F}{2r}\, \sigma_{AB}
\label{5.22}
\end{equation} 
for the nonvanishing components of the transverse curvature. 

The transverse curvature is discontinuous at $\Sigma$, and
Eqs.~(\ref{2.12})--(\ref{2.14}) allow us to compute the shell's
surface quantities. Because the generators are affinely parameterized 
by $-r$ on both sides of the hypersurface, we have that $p = 0$ ---
the shell has a vanishing surface pressure. On the other hand, its
surface density is given by 
\begin{equation}
\mu = \frac{m}{4\pi r^2},
\label{5.23}
\end{equation} 
the ratio of the shell's gravitational mass $m$ to its (ever
decreasing) surface area $4\pi r^2$. Thus far our results are
virtually identical to those obtained in the first subsection. What is
new in this context is the presence of a surface current $j^A$, whose
nonvanishing component is       
\begin{equation}
j^\psi = \frac{3Ma}{8\pi r^4}. 
\label{5.24}
\end{equation} 
This comes from the shell's rotation, and the fact that the situation
is not entirely spherically symmetric.  

To better understand the physical significance of the surface current,
we express the shell's surface stress-energy tensor,  
\begin{equation}
S^{\alpha\beta} = \mu k^\alpha k^\beta + j^\psi \bigl( k^\alpha
\base{\beta}{\psi} + \base{\alpha}{\psi} k^\beta \bigr), 
\label{5.25}
\end{equation} 
in terms of the vector $\ell^\alpha \equiv k^\alpha +
(j^\psi/\mu)\, \base{\alpha}{\psi}$. This vector is null (in the
slow-rotation limit) and has the components     
\begin{equation}  
\ell^\alpha \partial_\alpha = \frac{1}{f}\, \partial_t - \partial_r +
\frac{1}{f}\, \Omega_{\rm fluid} \partial_\phi
\label{5.26}
\end{equation} 
in the coordinates $x^\alpha = (t,r,\theta,\phi)$ used in 
${\cal M}^+$; we have set 
\begin{equation} 
\Omega_{\rm fluid} \equiv \frac{2Ma}{r^3} + \frac{3Ma}{2mr}\, f.  
\label{5.27}
\end{equation} 
The shell's surface stress-energy tensor is now given by the simple 
expression  
\begin{equation} 
S^{\alpha\beta} = \mu\, \ell^\alpha \ell^\beta, 
\label{5.28}
\end{equation} 
which corresponds to a pressureless fluid of density $\mu$ moving 
with a four-velocity $\ell^\alpha$. We see that the fluid is moving
along null curves (not geodesics!) that do not coincide with the
shell's null generators. This motion across generators is created by a
mismatch between $\Omega_{\rm fluid}$, the fluid's angular velocity,
and $\Omega_{\rm generators}$, the angular velocity of the
generators. The discrepancy is directly related to $j^A$:   
\begin{equation} 
\Omega_{\rm relative} \equiv \Omega_{\rm fluid} 
- \Omega_{\rm generators} = \frac{j^\psi}{\mu} 
= \frac{3Ma}{2mr}\, f.          
\label{5.29}
\end{equation} 
Notice that the fluid rotates faster than the generators, which share 
their angular velocity with zero-angular-momentum observers within
${\cal M}^-$. Notice also that $\Omega_{\rm relative}$ decreases to
zero as $r$ approaches $2M$: the fluid ends up corotating with the
generators when the shell crosses the black-hole horizon.  

The application considered in this subsection is a variation of an
analysis presented previously by Musgrave and Lake
\cite{musgrave2}. In some respects our discussion is less general than
theirs: our black hole is initially nonrotating and it does not carry
an electric charge. In other respects our discussion is more general:
here the shell's material is not restricted to move along the null 
generators; as we have seen, this generalization leads to interesting
phenomena.        

\subsection*{Cosmological phase transition}

In this fourth (and last) application of the formalism, we consider
an intriguing (but entirely artificial) cosmological scenario
according to which the universe was expanding initially in two
directions only, but was then made to expand isotropically by a sudden 
explosive event. This application is a simplified version of an
example previously presented by Barrab\`es and Hogan \cite{BH3}.  

The ${\cal M}^-$ region of spacetime is the one in which the universe
is expanding in the $x$ and $y$ directions only. Its metric is    
\begin{equation}
ds_-^2 = -dt^2 + a^2(t) \bigl( dx^2 + dy^2 \bigr) + dz_-^2,      
\label{5.30}
\end{equation} 
and the scale factor is assumed to be given by $a(t) \propto 
t^{1/2}$. The cosmological fluid moves with a four-velocity $u^\alpha 
= \partial x^\alpha/\partial t$, and it has a density and (isotropic)
pressure given by $\rho_- = p_- = 1/(32\pi t^2)$, respectively.   

In the ${\cal M}^+$ region of spacetime, the universe expands
uniformly in all three directions. Here the metric is 
\begin{equation}
ds_+^2 = -dt^2 + a^2(t) \bigl( dx^2 + dy^2 + dz_+^2 \bigr),   
\label{5.31}
\end{equation} 
with the same scale factor $a(t)$ as in ${\cal M}^-$, and the
cosmological fluid has a density and pressure given by $\rho_+ = 3 p_+
= 3/(32\pi t^2)$, respectively; this corresponds to a
radiation-dominated universe.  

The history of the explosive event that changes the metric from 
$g^-_{\alpha\beta}$ to $g^+_{\alpha\beta}$ coincides with a null
hypersurface $\Sigma$. This surface moves in the positive 
$z_\pm$ direction, and as we shall see, it supports a singular
stress-energy tensor. The ``agent'' that alters the course of the
universe's expansion is therefore a null shell.           

As seen from ${\cal M}^-$, the hypersurface is described by 
$t = z_- + \mbox{constant}$, and the vector $k^\alpha  
\partial_\alpha = \partial_t + \partial_z$ is tangent to 
the null generators, which are parameterized by $t$. Because   
$k^\alpha_{\ ;\beta} k^\beta = 0$, we have that $t$ is an affine  
parameter on this side of the hypersurface. The coordinates $x$ and
$y$ are constant on the generators, and we use them, together with
$t$, as intrinsic coordinates on $\Sigma$. We therefore have  
$y^a = (t,\theta^A)$, $\theta^A = (x,y)$, and the shell's induced
metric is    
\begin{equation} 
\sigma_{AB}\, d\theta^A d\theta^B = a^2(t)(dx^2 + dy^2). 
\label{5.32}
\end{equation} 
The remaining basis vectors are $\base{\alpha}{x} \partial_\alpha = 
\partial_x$, $\base{\alpha}{y} \partial_\alpha = \partial_y$, and
$N_\alpha\, dx^\alpha = -\frac{1}{2}(dt + dz_-)$. The nonvanishing  
components of the transverse curvature are 
\begin{equation} 
C^-_{AB} = \frac{1}{4t}\, \sigma_{AB}. 
\label{5.33}
\end{equation} 
We note that on the ${\cal M}^-$ side of $\Sigma$, the null generators 
have an expansion given by $\theta = k^\alpha_{\ ;\alpha} = 1/t$, and
that $T_{\alpha\beta} k^\alpha k^\beta = \rho_- + p_- 
= 1/(16\pi t^2)$, where $T^{\alpha\beta}$ is the stress-energy tensor
of the cosmological fluid.  

As seen from ${\cal M}^+$, the description of the hypersurface is
obtained by integrating $dt = a(t)\, dz_+$, and $k^\alpha 
\partial_\alpha = \partial_t + a^{-1} \partial_z$ is tangent to 
the null generators. We note that $t$ is {\it not} an affine parameter 
on this side of the hypersurface: we have that $k^\alpha_{\ ;\beta} 
k^\beta = (2t)^{-1} k^\alpha$. The remaining basis vectors are
$\base{\alpha}{x} \partial_\alpha = \partial_x$, $\base{\alpha}{y}
\partial_\alpha = \partial_y$, $N_\alpha\, dx^\alpha =
-\frac{1}{2}(dt + a\, dz_+)$, and the nonvanishing components of the  
transverse curvature are now   
\begin{equation} 
C^+_{tt} = \frac{1}{2t}, \qquad 
C^+_{AB} = \frac{1}{4t}\, \sigma_{AB}. 
\label{5.34}
\end{equation} 
On this side of $\Sigma$, the generators have an expansion also given  
by $\theta = 1/t$ (since continuity of $\theta$ is implied by
continuity of the induced metric), and $T_{\alpha\beta} k^\alpha 
k^\beta = \rho_+ + p_+ = 1/(8\pi t^2)$.  

The fact that $t$ is an affine parameter on one side of the
hypersurface only tells us that the shell must possess a surface
pressure. In fact, continuity of $C_{AB}$ across the hypersurface
implies that $p$ is the {\it only} nonvanishing surface quantity. It
is given by  
\begin{equation} 
p = -\frac{1}{16 \pi t}, 
\label{5.35}
\end{equation} 
the negative sign indicating that this surface quantity would be
better described as a {\it tension}, not a pressure. The shell's
surface stress-energy tensor is $S^{\alpha\beta} = p\, \sigma^{AB}\, 
\base{\alpha}{A} \base{\alpha}{B}$. If we select observers comoving
with the cosmological fluid as our preferred set of observers, then
$-k_\alpha u^\alpha = 1$ and the full stress-energy tensor of the
surface layer is given by $T_\Sigma^{\alpha\beta} = S^{\alpha\beta}
\delta(t - t_\Sigma)$, with $t_\Sigma$ denoting the time at which a
given observer crosses the shell. We see that for these observers,
$-p$ is the physically-measured surface tension.  

Finally, we note that the expressions $-p = 1/(16\pi t)$, $\theta =    
1/t$, and $[T_{\alpha\beta} k^\alpha k^\beta] = 1/(16\pi t^2)$ are 
compatible with the general relation $-p\, \theta = [T_{\alpha\beta}  
k^\alpha k^\beta]$ presented in Eq.~(\ref{4.5}). This shows that the
energy released by the shell as it expands is absorbed by the
cosmological fluid, whose density increases by a factor of
$\rho_+/\rho_- = 3$; this energy is provided by the shell's surface
tension.    
   
\begin{acknowledgments} 

This work was supported by the Natural Sciences and Engineering
Research Council of Canada. I am grateful to Claude Barrab\`es and
Werner Israel for their comments, and to Kayll Lake for his
encouragement to publish this work.  

\end{acknowledgments}

\end{document}